\documentclass[prl,twocolumn,superscriptaddress]{revtex4-2}

\usepackage{bm}
\usepackage{amsmath,amssymb,amsfonts}%
\usepackage{siunitx}
\usepackage{nicefrac}
\usepackage{braket}
\usepackage{graphicx}
\usepackage{soul}
\usepackage{xcolor}
\usepackage{ulem}

\usepackage[colorlinks = true]{hyperref}
\hypersetup{
     colorlinks   = true,
     citecolor    = blue,
     linkcolor    = blue,
     urlcolor     = blue
}

\begin{document}

\title[Article Title]{High-efficiency single-photon source above the loss-tolerant threshold for efficient linear optical quantum computing}

\author{Xing Ding}
\altaffiliation{These authors contributed equally to this work.}
\affiliation{Hefei National Research Center for Physical Sciences at the Microscale and School of Physical Sciences,
University of Science and Technology of China, Hefei, Anhui 230026, China}
\affiliation{CAS Center for Excellence in Quantum Information and Quantum Physics,
University of Science and Technology of China, Hefei, Anhui 230026, China}
\affiliation{Hefei National Laboratory,
University of Science and Technology of China, Hefei, Anhui 230088, China}

\author{Yong-Peng Guo}
\altaffiliation{These authors contributed equally to this work.}
\affiliation{Hefei National Research Center for Physical Sciences at the Microscale and School of Physical Sciences,
University of Science and Technology of China, Hefei, Anhui 230026, China}
\affiliation{CAS Center for Excellence in Quantum Information and Quantum Physics,
University of Science and Technology of China, Hefei, Anhui 230026, China}
\affiliation{Hefei National Laboratory,
University of Science and Technology of China, Hefei, Anhui 230088, China}

\author{Mo-Chi Xu}
\altaffiliation{These authors contributed equally to this work.}
\affiliation{Hefei National Research Center for Physical Sciences at the Microscale and School of Physical Sciences,
University of Science and Technology of China, Hefei, Anhui 230026, China}
\affiliation{CAS Center for Excellence in Quantum Information and Quantum Physics,
University of Science and Technology of China, Hefei, Anhui 230026, China}
\affiliation{Hefei National Laboratory,
University of Science and Technology of China, Hefei, Anhui 230088, China}

\author{Run-Ze Liu}
\altaffiliation{These authors contributed equally to this work.}
\affiliation{Hefei National Research Center for Physical Sciences at the Microscale and School of Physical Sciences,
University of Science and Technology of China, Hefei, Anhui 230026, China}
\affiliation{CAS Center for Excellence in Quantum Information and Quantum Physics,
University of Science and Technology of China, Hefei, Anhui 230026, China}
\affiliation{Hefei National Laboratory,
University of Science and Technology of China, Hefei, Anhui 230088, China}

\author{Geng-Yan Zou}
\affiliation{Hefei National Research Center for Physical Sciences at the Microscale and School of Physical Sciences,
University of Science and Technology of China, Hefei, Anhui 230026, China}
\affiliation{CAS Center for Excellence in Quantum Information and Quantum Physics,
University of Science and Technology of China, Hefei, Anhui 230026, China}
\affiliation{Hefei National Laboratory,
University of Science and Technology of China, Hefei, Anhui 230088, China}

\author{Jun-Yi Zhao}
\affiliation{Hefei National Research Center for Physical Sciences at the Microscale and School of Physical Sciences,
University of Science and Technology of China, Hefei, Anhui 230026, China}
\affiliation{CAS Center for Excellence in Quantum Information and Quantum Physics,
University of Science and Technology of China, Hefei, Anhui 230026, China}
\affiliation{Hefei National Laboratory,
University of Science and Technology of China, Hefei, Anhui 230088, China}

\author{Zhen-Xuan Ge}
\affiliation{Hefei National Research Center for Physical Sciences at the Microscale and School of Physical Sciences,
University of Science and Technology of China, Hefei, Anhui 230026, China}
\affiliation{CAS Center for Excellence in Quantum Information and Quantum Physics,
University of Science and Technology of China, Hefei, Anhui 230026, China}
\affiliation{Hefei National Laboratory,
University of Science and Technology of China, Hefei, Anhui 230088, China}

\author{Qi-Hang Zhang}
\affiliation{Hefei National Research Center for Physical Sciences at the Microscale and School of Physical Sciences,
University of Science and Technology of China, Hefei, Anhui 230026, China}
\affiliation{CAS Center for Excellence in Quantum Information and Quantum Physics,
University of Science and Technology of China, Hefei, Anhui 230026, China}
\affiliation{Hefei National Laboratory,
University of Science and Technology of China, Hefei, Anhui 230088, China}

\author{Hua-Liang Liu}
\affiliation{Hefei National Research Center for Physical Sciences at the Microscale and School of Physical Sciences,
University of Science and Technology of China, Hefei, Anhui 230026, China}
\affiliation{CAS Center for Excellence in Quantum Information and Quantum Physics,
University of Science and Technology of China, Hefei, Anhui 230026, China}
\affiliation{Hefei National Laboratory,
University of Science and Technology of China, Hefei, Anhui 230088, China}

\author{Lin-Jun Wang}
\affiliation{Center for Micro and Nanoscale Research and Fabrication,
University of Science and Technology of China, Hefei, Anhui 230026, China}

\author{Ming-Cheng Chen}
\affiliation{Hefei National Research Center for Physical Sciences at the Microscale and School of Physical Sciences,
University of Science and Technology of China, Hefei, Anhui 230026, China}
\affiliation{CAS Center for Excellence in Quantum Information and Quantum Physics,
University of Science and Technology of China, Hefei, Anhui 230026, China}
\affiliation{Hefei National Laboratory,
University of Science and Technology of China, Hefei, Anhui 230088, China}

\author{Hui Wang}
\affiliation{Hefei National Research Center for Physical Sciences at the Microscale and School of Physical Sciences,
University of Science and Technology of China, Hefei, Anhui 230026, China}
\affiliation{CAS Center for Excellence in Quantum Information and Quantum Physics,
University of Science and Technology of China, Hefei, Anhui 230026, China}
\affiliation{Hefei National Laboratory,
University of Science and Technology of China, Hefei, Anhui 230088, China}

\author{Yu-Ming He}
\affiliation{Hefei National Research Center for Physical Sciences at the Microscale and School of Physical Sciences,
University of Science and Technology of China, Hefei, Anhui 230026, China}
\affiliation{CAS Center for Excellence in Quantum Information and Quantum Physics,
University of Science and Technology of China, Hefei, Anhui 230026, China}
\affiliation{Hefei National Laboratory,
University of Science and Technology of China, Hefei, Anhui 230088, China}

\author{Yong-Heng Huo}
\affiliation{Hefei National Research Center for Physical Sciences at the Microscale and School of Physical Sciences,
University of Science and Technology of China, Hefei, Anhui 230026, China}
\affiliation{CAS Center for Excellence in Quantum Information and Quantum Physics,
University of Science and Technology of China, Hefei, Anhui 230026, China}
\affiliation{Hefei National Laboratory,
University of Science and Technology of China, Hefei, Anhui 230088, China}

\author{Chao-Yang Lu}
\affiliation{Hefei National Research Center for Physical Sciences at the Microscale and School of Physical Sciences,
University of Science and Technology of China, Hefei, Anhui 230026, China}
\affiliation{CAS Center for Excellence in Quantum Information and Quantum Physics,
University of Science and Technology of China, Hefei, Anhui 230026, China}
\affiliation{Hefei National Laboratory,
University of Science and Technology of China, Hefei, Anhui 230088, China}

\author{Jian-Wei Pan}
\affiliation{Hefei National Research Center for Physical Sciences at the Microscale and School of Physical Sciences,
University of Science and Technology of China, Hefei, Anhui 230026, China}
\affiliation{CAS Center for Excellence in Quantum Information and Quantum Physics,
University of Science and Technology of China, Hefei, Anhui 230026, China}
\affiliation{Hefei National Laboratory,
University of Science and Technology of China, Hefei, Anhui 230088, China}

\date{\today}

\begin{abstract}
Photon loss is the biggest enemy for scalable photonic quantum information processing. This problem can be tackled by using quantum error correction, provided that the overall photon loss is below a threshold of 1/3. However, all reported on-demand and indistinguishable single-photon sources still fall short of this threshold. Here, by using tailor shaped laser pulse excitation on a high-quantum efficiency single quantum dot deterministically coupled to a tunable open microcavity, we demonstrate a high-performance source with a single-photon purity of 0.9795(6), photon indistinguishability of 0.9856(13), and an overall system efficiency of 0.712(18), simultaneously. This source for the first time reaches the efficiency threshold for scalable photonic quantum computing. With this source, we further demonstrate 1.89(14) dB intensity squeezing, and consecutive 40-photon events with 1.67 mHz count rate.
\end{abstract}

\maketitle

Quantum computers can solve certain hard problems beyond the reach of any classical computer. Recent noisy intermediate-scale quantum (NISQ) processors implementing Gaussian boson sampling with photons \cite{zhong2020quantum} or random circuit sampling with superconducting qubits \cite{arute2019quantum} have poised for the technological singularity called quantum computational advantage, which raised serious challenges to the extended Church-Turing Thesis \cite{bernstein1993quantum}. However, without effective quantum error correction, the NISQ processing fidelity decreases exponentially when scaling up the size. Building fault-tolerant universal quantum computers represents the most important challenge ahead \cite{bharti2022noisy}. 

Single photons are fast-flying, can be operated at room temperature, and have very weak interactions with the environment. However, the single photons can be easily lost either owing to inefficient photon sources and detectors, or in the lossy photonic circuits. This represents arguably the biggest enemy of universal photonic quantum computing.

The celebrated threshold theorem \cite{aharonov1997fault} states that if the physical error rate is below a certain threshold, arbitrarily long calculations at arbitrarily low error rates, i.e., fault-tolerant quantum computing, are possible through quantum error correction. To tackle the problem of photon loss, various quantum error correction codes have been developed \cite{grassl1997codes,wasilewski2007protecting}. In particular, in the measurement-based model of quantum computation, Varnava, Browne, and Rudolph \cite{varnava2008good} showed that, if the product of single-photon source and detection efficiency is above 2/3, then linear optical quantum computing is possible. Such an extremely relaxed photon loss threshold for the primary error mechanisms in the photonic system has motivated extensive efforts in developing increasingly high-efficiency quantum light sources.

Over the past 50 years, tremendous progress has been made in the performance of the quantum light source, from the pioneering entangled photons used for Bell test \cite{aspect1981experimental} to the quantum computing experiments recently with tens of indistinguishable photons \cite{zhong2020quantum,zhong2021phase, madsen2022quantum, wang2019boson}. In particular, spontaneous parametric down-conversion or four-wave mixing can provide heralded single photons based on the probabilistically generated two-photon pairs, which have been engineered to have simultaneously near-unity collection efficiency and indistinguishability \cite{zhong201812,paesani2020near}. To overcome its probabilistic, spatial or temporal multiplexing has been attempted, however, the best source efficiency achieved at a cost of high multi-photon noise ($g^2(0)=0.27$) still cannot exceed the threshold \cite{kaneda2019high}. 

Quantum dots (QDs) are in principle deterministic single-photon emitters. Pulsed resonant excitation to the single QDs embedded in finely tuned microcavities such as micropillars \cite{ding2016demand, wang2019towards}, bullseye \cite{wang2019towards}, and open cavity \cite{tomm2021bright} have created single photons with increasingly high efficiency and indistinguishability. However, the best reported end-to-end single-photon source efficiency is 0.57 \cite{tomm2021bright}, still below the loss-tolerant threshold. In this letter, we report a single-photon source, for the first time, reaches the loss tolerant efficiency threshold for scalable photonic quantum computing, with a system efficiency of 0.712(18), while maintaining a single-photon purity of 0.9795(6) and indistinguishability of 0.9856(13). 

\begin{figure}
\includegraphics[width=\columnwidth]{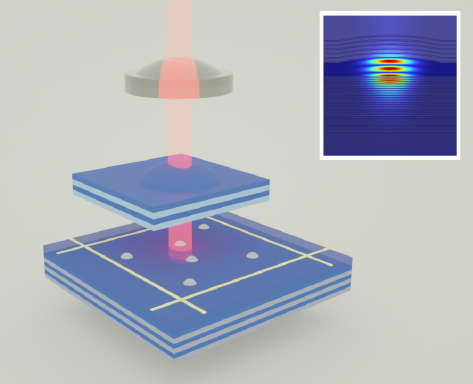}
\caption{The design of the QDs in a fully tunable open-cavity setup. The top concave cavity mirror is mounted on a stack of three piezo nano-positioners. The bottom part, including 30 pairs DBR and QD layer, is mounted on a stack of three nano-positioners (XYZ) and two tilt stages ($\Theta, \Phi$). The $\lambda$-thickness InAs QD layer is grown on top of a 30 pairs AlAs/GaAs DBR. The center wavelength of the bottom DBR is 890 nm. The top surface of the QD layer is coated with 50 nm thickness Au/Ti grids, which facilitates the location of the selected QDs with high quantum efficiency. This setup provides the in-situ ability to align the cavity with selected QDs. The inset shows a numerical simulation of the radiation from a dipole in an open-cavity structure. The numerically simulated $Q$-factor is $\sim$9000 and the Purcell factor is 18. The probability that single photons can transmit through the top of the cavity is $\eta_{top}=0.939$.}
\label{fig1}
\end{figure}

To optimize the single QD-microcavity coupling, we use a tunable open cavity \cite{tomm2021bright}, as shown in Fig.~\ref{fig1}(a), in a plane-concave Fabry-Perot cavity configuration. The advantage of the open microcavity is that it provides full degrees of freedom to maximize the coupling between selected QDs with high quantum efficiencies and the microcavity spectrally and spatially. The open cavity consists of a 5.5 pairs SiO${}_2$/Ta${}_2$O${}_5$ concave distributed Bragg reflector (DBR) layer in the top, and in the bottom a $\lambda$-thickness QD membrane, which is directly grown on the 30 pairs AlAs/GaAs planar DBR layer by molecular beam epitaxy. The top cavity mirror is fabricated using FIB etching concave surface on a SiO${}_2$ substrate. The radius of curvature and diameter of the concave mirror is 15 $\mu$m and 6.3 $\mu$m respectively. The reflectivity of the top mirror is $\sim$98.6\%, while the reflectivity of the bottom mirror is $\sim$99.97\%.

The QD-open microcavity system is cooled down to 4 K in a closed-cycle cryostat. Note that while the open cavity brings the key advantage of full tunability, it comes with an extreme technical challenge of high stability, because even a 1 nm drift of the cavity length would cause a shift of the cavity's resonance frequency and a drop of 78\% of the maximum Purcell factor. We made various efforts to overcome the environmental vibrations, especially that from the cold head. The cryostat is mounted on a passive vibration damper to isolate vibrations from the environment. 

\begin{figure*}[ht]%
\centering
\includegraphics[width=1.0\textwidth]{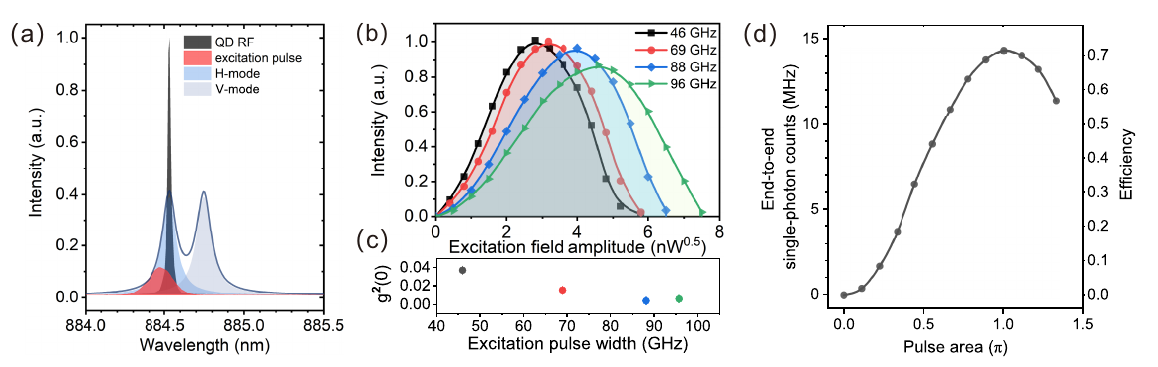}
\caption{High-efficiency single-photon source (SPS) under shape-optimized pulse excitation. (a) Excitation scheme of a polarized single-photon source in a mode-splitting cavity. (b) Deterministic generation of polarized single photons under a series of pulse excitations with different pulse widths. (c) The corresponding single-photon purity under different pulse width excitations. (d) The RF photon counts as a function of the square root of excitation laser power when the pulse width is 69 GHz. At the $\pi$ pulse, the end-to-end single-photon counts per second reach a maximum of $(14.28\pm0.01)$ million, under 25.38 MHz repetition rate laser excitation, which corresponds to a system efficiency of 0.712(18).}\label{fig2}
\end{figure*}

To maximize the brightness of the single-photon emitters, we first pre-select QDs with high quantum efficiencies by removing the top concave cavity and comparing the $\pi$-pulse resonance fluorescence (RF) of hundreds of bare QDs. The positions of the brightest candidates are recorded. Then, the top mirror returns, forming a fully tunable open microcavity that allows in-situ coupling of the cavity spatially and spectrally. 

The basic excitation protocol follows the resonant excitation of a QD in a degenerate mode-splitting cavity \cite{wang2019towards}. The mode-splitting here is induced by the birefringence of the GaAs material, and the magnitude of mode-splitting can be manipulated by applying strains in the GaAs material \cite{tomm2021tuning}. Compared with geometrically asymmetric cavities, birefringence-induced mode-splitting does not induce far-field optical mode deformation, which is favorable for higher single-mode fiber collection efficiency. Figure~\ref{fig2}(a) shows the experimental excitation scheme used in practice. The measured $Q$-factor of each mode is $\sim$8400 and splitting between two modes is $\sim$83 GHz, which gives the ratio of the cavity splitting ($\Delta \omega$) to the cavity linewidth ($\delta \omega$) of 2.07. A singly negatively charged QD emitting at a wavelength of 884.5 nm with high quantum efficiency $\eta_{qd}$ is coupled to the cavity's H-polarized mode. Then we can excite the QD with a V-polarized laser pulse and collect the H-polarized single photons.

It is important to note that, in previous experiment \cite{wang2019towards}, due to the frequency mismatch between the V-polarized excitation laser and the H-polarized cavity mode, only a small fraction of the laser can actually be used to excite the QD. The frequency of the partial excitation laser may be far detuned with the QD, leading to a reduction of the $\pi$ pulse single-photon counts. Compared to low $Q$ resonator structures like bullseye cavity, this problem is more severe in high $Q$ resonators such as micro-pillar and open-cavity.

To overcome this problem, we use a 4f optical pulse shaping system to shape the laser pulse. The shaping system consists of two reflective diffraction gratings, two identical lenses, and one slit. The groove density of the gratings is 1200g/mm, and the focal length of the lenses is 1830 mm. The first grating is used to transform frequency components into spatial ones, and the second grating is for an inverse process. According to the detailed description of the excitation scheme in supplementary material \cite{sm}, with a narrower excitation pulse width in frequency, the intra-cavity pulse would resonantly excite the QD more effectively. Thus we set a width-tunable slit at the Fourier plane to filter out unwanted frequency components. As shown in fig~\ref{fig2}(b), by resonantly exciting the selected QD with different pulse widths, from $\sim$96 GHz to $\sim$46 GHz, a set of clear Rabi oscillations are observed. These oscillations illustrate an improvement in the $\pi$ pulse single-photon counts, as well as a reduction in the power required for their excitation when utilizing narrower pulse widths.

It should also be noted that narrowing the pulse width in frequency will broaden the pulse length in the time domain, which can cause re-excitation of the QD and reduce the single-photon purity. In fig~\ref{fig2}(c), we show the measured second-order correlations of the single photons under different excitation pulse widths. By comparing the purity and brightness of the single photons, we can figure out that 69 GHz is the most optimized pulse width to excite our selected QD in the open cavity. Under this shape-optimized pulse excitation, a single-photon source with minimum excitation power, near unity single-photon purity, and maximum single-photon counts can be achieved simultaneously. The shaped pulse excitation generates 13\% more single photons at the $\pi$ pulse, in comparison to the unmodified pulse excitation with a pulse width of 96 GHz.

Having optimized the shaping system, we proceeded to measure the efficiency of our single-photon system. To avoid the dead time zone (30ns) of the superconducting nanowire single-photon detector (SNSPD), we utilized an amplitude electro-optic modulator to decrease the repetition rate of the excitation laser to 25.38 MHz, which is one-third of the original frequency of 76.13 MHz. The single photons collected at the single-mode fiber end are directly connected to the SNSPD, whose detection efficiency is 0.79(2). In Figure~\ref{fig2}(d), we present the corrected RF photon counts plotted against the square root of the excitation laser power. Using a $\pi$ pulse power of 10.24 nW, we observed a $\pi$ pulse single-photon count rate of approximately 14.28 million counts per second, resulting in a system efficiency of 0.712(18). This is the highest achieved single-photon source's system efficiency of all reported physical systems \cite{sm}. 

\begin{figure*}[ht]%
\centering
\includegraphics[width=1.0\textwidth]{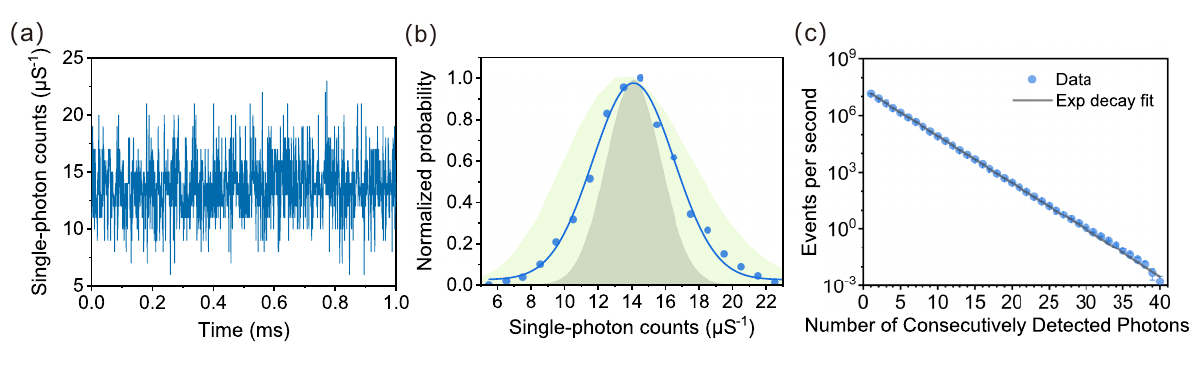}
\caption{System efficiency characterization of the single-photon source. (a) The single-photon stream with a time bin of 1.0us. 1000 bins of single-photon counts are recorded. (b) The corresponding normalized histogram. The intensity fluctuation of single photons is shown as deep blue dots and fitted with a binomial function. The green shadow shows the shot-noise-limited source with the same photon count rate. The directly measured intensity squeezing of the single-photon source compared with the shot-noise limit is $\sigma_{SPS}/\sigma_{SN}=0.65(2)$ under $\pi$ pulse excitation. The grey shadow shows the corrected intensity squeezing value at the first lens is $3.92\pm0.18$ dB. (c) Statistics of consecutive n-photon events within 10 minutes, up to 40 photons in a row can be acquired, the line indicates an exponential decay fit.
}\label{fig3}
\end{figure*}

An important feature of high-efficiency single-photon source can be manifested by intensity squeezing \cite{mandel1979sub}, i.e., the reduction of intensity fluctuation. Such an observation was only possible very recently \cite{wang2020observation}, after 20 years development of QD single photons, highlighting the importance of improving the system efficiency of single-photon sources. The standard deviation of counts in a single-photon source is reduced by an intensity squeezing factor of $\sqrt{(1-\rho)}$ compared to the shot noise, so the value of intensity squeezing is a natural benchmark to characterize the overall efficiency of single-photon sources. Here $\rho$ is the overall efficiency, including both the collection efficiency and the detection efficiency. A time-correlated single-photon counting (TCSPC) system is used to record the arrival time of each photon for further analysis. Figure~\ref{fig3}(a) shows the real-time detected single photons at $\pi$ pulse with a time bin of 1.0 $\mu$s, the corresponding histogram is shown in Fig~\ref{fig3}(b). The standard deviation of single-photon counts (blue) in Fig~\ref{fig3}(b) is 2.43(8), which shows sub-shot-noise intensity fluctuation with an intensity squeezing of 1.89(14) dB. 

We analyzed the consecutive single-photon events to reveal the overall efficiency. Figure~\ref{fig3}(c) illustrates the number of consecutively detected n-photon streams. The integration time is 10 minutes, and a maximum of 40-photon events are observed at a count rate of 1.67 mHz. The event rate of consecutive $n$-photons should be $\rho^{n} \times R_{laser}$, here $R_{laser}$ is the repetition rate of the pulse laser. By fitting the exponential decay, the overall efficiency $\rho$ was determined to be $0.5652(7)$. Taking into account the detection efficiency of the used detector is $0.79(2)$, the system efficiency of the single-photon source is $0.715(18)$, which again confirms the efficiency measurement.

\begin{figure*}[ht]%
\centering
\includegraphics[width=1.0\textwidth]{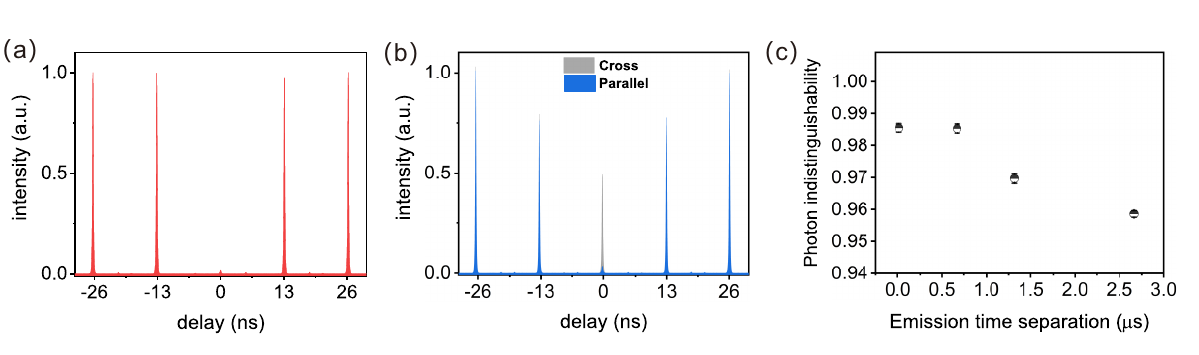}
\caption{Characterization of the pulsed RF single photons. (a) The intensity correlation histogram of the single photons under $\pi$ pulse excitation measured with an HBT setup. The second-order correlation at zero time delay is $g^2(0)=0.0205(6)$. (b) Quantum interference between two single photons. The input two photons are $\pi$ pulse excited and prepared in cross and parallel polarization. The raw HOM visibility is 0.928(1). (c) Single-photon indistinguishability as a function of emission time separation. The corrected photon indistinguishability drops very slightly, from 0.9856(13) at 13.1 ns to 0.959(2) at $\sim$2.67 $\mu$s.
}\label{fig4}
\end{figure*}

In addition to measuring system efficiency, we conducted comprehensive characterizations of the single-photon source by evaluating its purity and indistinguishability. The single-photon purity is characterized with a Hanbury Brown and Twiss (HBT) setup \cite{hanbury1979test} and presented in Fig~\ref{fig4}(a), with $g^2(0)=0.0205(6)$. The data are directly measured without any further filtering. The excitation laser is found to be the primary source of residual multi-photon events. The photons' indistinguishability is measured using a non-postselective two-photon Hong-Ou-Mandel interferometer (HOM) \cite{hong1987measurement}. Figure~\ref{fig4}(b) shows the histograms of the normalized two-photon counts for cross and parallel polarization respectively. The raw two-photon quantum interference visibility is 0.928(1). After correction with the residual multi-photon probability of $g^2(0)=0.0205(6)$ and an unbalanced beam splitter split ratio of 45:55, the corrected single-photon indistinguishability is 0.9856(13).

To further prove that the single-photon source is ready for scale-up, it is imperative to verify that that the single photons are nearly transform-limited \cite{wang2016near,kuhlmann2015transform}. Here we test the photon indistinguishability with different photon time delays using a delay line adjustable HOM interferometer \cite{wang2016near}. As shown in Fig~\ref{fig4}(c), the corrected HOM visibility is 0.9856(13), 0.985(1), 0.970(2), and 0.959(1) for the time delays of 13.1 ns, 670 ns, 1.31 $\mu$s and 2.67 $\mu$s, respectively. Compared with the 13.1 ns delay, the single-photon indistinguishability with 2.67 $\mu$s delay only drops less than 3\%, indicating that the emitted single photons are close to the transform limit. Such high photon indistinguishability with long-time delay guarantees the performance in scalable photonic applications.

In conclusion, by resonant excitation of a QD embedded in an open-cavity, we have realized a deterministic single-photon source with single-photon purity of 0.9795(6), indistinguishability of 0.9856(13), and system efficiency of 0.712(18) simultaneously, which for the first time exceeds the required efficiency threshold for scalable photonic quantum computing \cite{varnava2008good}. Such a high-efficiency source can be immediately used for applications such as boson sampling, photonic cluster states generation, quantum communication, etc. Furthermore, this QD-open cavity system can be readily extended to the strong coupling regime by increasing the coupling rate, thereby facilitating the implementation of photon-photon quantum gates. With its exceptional performance and significant potential for various applications, our work marks a critical step towards scalable quantum technologies reliant on single photons.

\begin{acknowledgements}
Our work is supported by the National Natural Science Foundation of China (Grant No.\ 12012422), the National Key R\&D Program of China (Grant No.\ 2019YFA0308700), the Chinese Academy of Sciences, the Anhui Initiative in Quantum Information Technologies (Grant No.\ AHY060000), the Science and Technology Commission of Shanghai Municipality (Grant No.\ 2019SHZDZX01), the Innovation Program for Quantum Science and Technology (Grant No.\ 2021ZD0301400, 2021ZD0300204) and China Postdoctoral Science Foundation (Grant No.\ 2021M703102).
\end{acknowledgements}

\normalem
\bibliographystyle{apsrev4-2}
\bibliography{sn-article}

\end{document}